# Optimal strengthening of particle-loaded liquid foams

F. Gorlier[1], Y. Khidas[2], A. Fall[1] and O. Pitois[1,*]

[1] Université Paris Est, Laboratoire Navier, UMR 8205 CNRS – École des Ponts ParisTech – IFSTTAR
cité Descartes, 2 allée Kepler, 77420 Champs-sur-Marne, France.
francois.gorlier@ifsttar.fr ; abdoulaye.fall@ifsttar.fr ; olivier.pitois@ifsttar.fr

[2] Université Paris Est, Laboratoire Navier, UMR 8205 CNRS – École des Ponts ParisTech – IFSTTAR
5 bd Descartes, 77454 Marne-la-Vallée Cedex 2, France.
yacine.khidas@u-pem.fr

* corresponding author

Abstract:

Foams made of complex fluids such as particle suspensions have a great potential for the development of advanced aerated materials. In this paper we study the rheological behavior of liquid foams loaded with granular suspensions. We focus on the effect of small particles, i.e. particle-to-bubble size ratio smaller than 0.1, and we measure the complex modulus as a function of particle size and particle volume fraction. With respect to previous work, the results highlight a new elastic regime characterized by unequaled modulus values as well as independence of size ratio. A careful investigation of the material microstructure reveals that particles organize through the network between the gas bubbles and form a granular skeleton structure with tightly packed particles. The latter is proven to be responsible for the reported new elastic regime. Rheological probing performed by strain sweep reveals a two-step yielding of the material: the first one occurs at small strain and is clearly attributed to yielding of the granular skeleton; the second one corresponds to the yielding of the bubble assembly, as observed for particle-free foams. Moreover the elastic modulus measured at small strain is quantitatively described by models for solid foams in assuming that the granular skeleton possesses a bulk elastic modulus of order 100 kPa. Additional rheology experiments performed on the bulk granular material indicate that this surprisingly high value can be understood as soon as the magnitude of the confinement pressure exerted by foam bubbles on packed grains is considered.

# 1. Introduction

Foaming of complex fluids is widely encountered in industrial processes as the first step in the elaboration of aerated materials. Such an aeration process is performed in order to improve thermal and acoustical performances of materials or simply to make them lighter and to save raw resources. Typical examples can be found in the production of materials for the building industry [1], of ceramic foams [2], of products for food [3] and cosmetic industries. Note also that the mining industry extensively resorts to mixtures of foam and particles through the flotation process that is widely used to separate ores [4].

The production of advanced aerated material requires to control the final pore morphology. For example, pore size and pore connectivity have strong effects on sound absorbing efficiency [5]. Unfortunately, several foam aging processes act before the complete hardening of foamy systems to damage the microstructure that has been set up during the aeration step: the drainage of the interstitial phase (the continuous phase between the gas bubbles) and the simultaneous rising of the bubbles, the ripening (the gas exchange between bubbles) and the coalescence process [6,7]. Recent studies have shown that those aging processes can be controlled in taking advantage of the rheological properties of fluids to be foamed, such as clays [8,9], coal fly ashes [10], colloidal suspensions [11–14], emulsions [15–17], granular suspensions [18,19]. In the elaboration of optimized foamed materials, the rheological behavior of those complex foams is also of primary interest. For example, elasticity and yield stress properties allow the material to sustain external forces and to keep its initial shape under gravity [20–22]. In contrast, filling of molds requires an appropriate workability of the fresh foamy paste. In this regard, rheology of aqueous foams [7,23] and rheology of suspensions [24,25] have been studied mostly separately and only few recent studies have however tackled the issue of coupling of interstitial rheology and bubble elasticity in foams [9,26]. Whereas a satisfactory modeling of the global rheology is still lacking, unusual and interesting rheological behaviors have been reported. Cohen-Addad et al. [26] provided a very enlightening example of the strong synergistic effects that can be brought about in complex liquid foams: it is shown that adding a tiny amount of non-colloidal particles in aqueous foams allows for elastic and loss shear moduli to be enhanced by more than one order of magnitude. Particle-size dependence is reported, i.e. the effect is all the more pronounced as the particle size is smaller, but however the authors do not conclude on strengthening behavior as the particle-to-bubble size ratio is further decreased. The investigation of such a behavior is precisely the purpose of this article. We therefore focus on the effect of small particle-to-bubble size ratios, i.e. smaller than 0.1, and we measure the complex modulus as a function of particle size and particle volume fraction. As shown in the following our results highlight a new elastic regime characterized by unequaled modulus values as well as independence of size ratio.

## 2. Materials and methods

Particle-loaded foams are prepared by mixing aqueous foam and granular suspension. The first step of the preparation is the production of precursor aqueous foam with well-controlled bubble size $D_b$ and gas volume fraction. Foaming liquid and perfluorohexane-saturated nitrogen are pushed through a T-junction allowing the control of the bubble size by adjusting the flow rate of each fluid; for the present study only one bubble size has been used: $D_b = 450$µm. Produced bubbles are collected in a glass column and constant gas fraction equal to 0.99 is set over the foam column by imbibition from the top with foaming solution. Secondarily, we prepare suspensions of polystyrene beads, with a volume fraction chosen within the range 0.05-0.55 and a monodisperse particle size $d_p \in [6, 10, 20, 30]$ µm. Note that those two components share the same continuous phase, which is composed of distilled water 80% w/w and glycerol 20% w/w, and trimethyl(tetradecyl)azanium bromide at a concentration 5 g.L$^{-1}$. Shear viscosity and density of that solution have been measured to be equal to 1.7 mPa.s and 1050 kg.m$^{-3}$ respectively. The surface tension of the solution has been measured to be equal to 38 mN.m$^{-1}$. The density of polystyrene beads is 1050 kg.m$^{-3}$ so that the latter are not subjected to sedimentation in the suspending liquid. Finally, precursor foam and granular suspension are mixed in a continuous process thanks to a mixing device based on flow-focusing method [19,27]. By tuning the flow rates of both the foam and the suspension during the mixing step, the gas volume fraction $\phi_0$ and the particle volume fraction $\phi_{p0}$ can be tuned. Note also that the bubble size is conserved during the mixing step, i.e. $D_b = 450$µm. Resulting particle-loaded foams are continuously poured into the measurement cell (cup geometry: height = 7 cm and diameter = 37mm). After this filling step, a six-bladed vane tool (height = 6 cm and diameter = 25mm) is inserted into the foam cell and the evolution of the sample is followed by measuring the shear elastic modulus through oscillatory rheometry (stress-controlled rheometer Malvern kinexus ultra+) with a strain of 10$^{-3}$ at 1Hz (see figure 1a). After a transient regime, the elastic modulus of all samples was found to reach a constant value. This behavior is attributed to gravity drainage, during which both liquid and particles can flow down through the bubble assembly. It has been shown in a previous work [19] that this flow, as well as the final equilibrium state, i.e. the final gas fraction $\phi$ and the final particle volume fraction $\phi_p$, are governed by the initial gas fraction $\phi_0$, the particle-to-bubble size ratio $d_p/D_b$ and the initial particle volume fraction within the interstitial suspension $\varphi_{p0} = \phi_{p0}/(1 - \phi_0)$. Therefore, for each particle size, a significant number of samples with different parameters $\phi_0$ and $\varphi_{p0}$ have been prepared in order to obtain drained samples with different parameters $\phi$ and $\phi_p$. Those two parameters are measured thanks to a second cell (height = 7 cm and diameter = 26mm) also filled during the generation step. The bottom of this cell is a piston allowing for the particle-loaded foam to be partially pushed out after drainage. This setup allows sampling the foam along its height and the particle fraction profile is measured as follows: each sampled volume is first weighted and then rinsed several times with ethanol in order to break the foam and to remove glycerol (centrifugation is performed for separating the particles from the liquid). Finally, the collected particle/ethanol mixture is let for drying (twelve hours at 60°C) and the resulting dried particles are weighted. Examples of vertical profiles are presented in figure 1b, showing that the filling/drainage procedure produces samples with very good homogeneity for high particle loadings, i.e. $\langle \phi_p \rangle \geq 0.02$, and reasonably homogeneous for the low particles loadings, where $0.7 \leq \Delta \phi_p / \langle \phi_p \rangle \leq 1.3$ for $\langle \phi_p \rangle \leq 0.01$. Such a behavior can be explained by the effect of initial particle concentration on drainage [19]. For high concentration, particles are trapped by collective jamming, i.e. channel size is larger than particle size. This situation promotes uniform concentration profile. For low concentration, collective jamming if rather ineffective and particle trapping is mainly due to individual captures, i.e. channel size is equal to particle size. For such cases, the quantity of trapped particles is strongly related to channel size at equilibrium, which is known to increase near the bottom: capture is less effective in those areas.

After the drainage step, the rheology measurement procedure starts: elastic and loss moduli are measured as a function of increased shear strain amplitude at frequency equal to 1Hz. Note that (i) to avoid slippage on the cell wall as shear stress is applied, the cell surface has been striated to jam the bubbles; (ii) the minimal gap in the vane-cup geometry represents 12 bubble diameters; (iii) the presence of perfluorohexane inside the bubbles strongly reduces the foam ripening rate which allows aging affects to be ignored over the time scale of measurements [28].

In addition to rheometry performed on particle-loaded foams, the rheology of the dry granular medium is also studied. In this case, polystyrene beads (diameter 500 µm) are poured into a cup geometry (diameter: 50 mm and bed depth: 8 mm). Then, a constant normal stress $P_C$ is exerted on the grains thanks to the controlled normal (vertical) force of the rheometer and using a plate-in-cup geometry (see Fiscina et al. [29] for more details). Since the grains cannot escape from the cell, the packing fraction can be measured from the gap variation: it varies from 0.625 to 0.628 for the applied confinement pressures. To avoid wall slip, both the moving upper boundary and the static lower boundary are serrated with 0.5mm ridges, which corresponds to the size of the grains. Then strain amplitude is increased while frequency is held constant (1Hz). Note that the maximum pressure exerted by the weight of grains in the layer is about 80 Pa, therefore the chosen values for the confinement pressure are such that $P_C \gg 80$ Pa.

## 3. Results

Results obtained for the elastic modulus $G'$ of foams loaded with 20 µm particles are presented as a function of strain amplitude $\varepsilon$ on figure 2a for several values of particle volume fraction $\phi_p$. $G'(\varepsilon \to 0)$ increases strongly as $\phi_p$ increases within the range 0-6%. Note that those values are by far larger than the one for particle-free foam, so that this strong increase is not expected to result from direct capillary effects. The critical deformation $\varepsilon_c$ marking the end of the linear regime decreases strongly as $\phi_p$ increases (see figure 2b) so that $G'$ decreases significantly at intermediate and higher deformations. Again, this behavior is different from the behavior of the particle-free foam, for which decrease in the $G'$ values is observed only for $\varepsilon > 10^{-1}$. In fact, $G'(\varepsilon)$ curves clearly show a first decrease for $10^{-3} < \varepsilon < 3\ 10^{-1}$, which can be attributed to the presence of the particles, and a second decrease for $\varepsilon > 3\ 10^{-1}$, which corresponds to the behavior classically observed for aqueous foams. Results for the loss modulus are presented in figure 2c. Similarly to $G'$, (1) $G''$ increases strongly at small strains, accounting for effects of particle loading, and (2) the behavior of aqueous foam disappears progressively as $\phi_p$ increases, except for $\varepsilon > 3\ 10^{-1}$. Note also that the slope $dG''/d\varepsilon$ observed at small strains becomes negative for $\phi_p > 0.03$.

Results for $G' \equiv G'(\varepsilon \to 0)$ are plotted in figure 3 as a function of $\phi_p$ for the four investigated particle sizes. Note that modulus values are divided by the modulus of the corresponding particle-free foam, i.e. $G'(\phi, 0) \approx b(\gamma/D_b)\phi(\phi - \phi_c)$ where $b \approx 2.8$ and $\phi_c \approx 0.64$ is the volume fraction corresponding to the random close packing of spheres [7]. Data obtained for $d_p \leq 20$ µm collapse on a single curve showing the strong increase of the elastic modulus with $\phi_p$. The reduced modulus is increased by a factor 25 as $\phi_p$ increases up to 6%. In contrast to previous work [26] there is no size-dependence effect within the investigated range of particle sizes. For $d_p = 30$ µm, the measured values for the elastic modulus are slightly smaller than for $d_p \leq 20$ µm.

Images of the foam microstructure are presented in figure 4, revealing the organization of particles within the bubble assembly. For small $\phi_p$ values the particles form aggregates that are mainly localized at nodes of the interstitial foam network. Increasing $\phi_p$ makes those aggregates to grow by invading the so-called Plateau borders connecting the nodes (4a), until the network be completely filled in the form of particle strings. From that $\phi_p$ value, the interstitial network form a skeleton made of particles, and the skeleton elements keep on enlarging as $\phi_p$ increases (4b).

We now turn to the results for the bulk granular material. Figure 5a shows the storage modulus as a function of strain amplitude for several values of confinement pressure $P_C$. The global behavior $G'(\varepsilon)$ depends significantly on $P_C$: at small confinement pressure $G'$ decreases starting from the smallest investigated strain values, which means that there is no linear regime for such strain values, whereas for larger confinement pressures a well-defined linear regime is observed and identifies the solid-like behavior of the material. Note that $G'$ is larger than $G''$ (see inset in figure 5). The yielding behavior of such granular material can be ascribed completely to the frictional properties of the grains in contact under the applied normal stresses. It is important to notice that for such granular matter, moderate confinement pressures induce significant values for the shear elastic modulus, as highlighted in figure 5b, where the ratio $G'/P_C$ is shown to range between $10^2$ and $3\ 10^3$ for investigated values of $P_C$.

## 4. Discussion

As a starting point, we compare our results with those obtained by Cohen-Addad et al. [26]. Note that we didn't use the same sizes for both particles and bubbles, and we therefore refer to the particle-to-bubble size ratio in the following. The authors analyze their results in terms of rigidity percolation threshold: they assume that the presence of capillary bridges, whose typical range is denoted $2h$, allows for the transmission of force necessary for rigidity percolation. According to the authors, the particle-laden foam could be compared to a dispersion of effective rigid elements of enhanced diameter $d_p + 2h$ that occupy a volume fraction $\phi_p^{eff} = \phi_p(1 + 2h/d_p)^3$. Under these conditions, the threshold $\Phi_{pc}$ is reached when a percolating cluster of effective rigid elements is formed. In the limit of small particle volume fractions, the threshold $\Phi_{pc}$ can be determined by fitting the relation $G'(\phi_p)/G'(0) \approx 1/(1 - \phi_p/\Phi_{pc})$ on the data [26]. We follow this approach to determine this percolation threshold $\Phi_{pc}$ (see the inset in figure 6) and we plot the corresponding values as a function of $d_p/D_b$ in figure 6. It appears that for the small $d_p/D_b$ values studied here, i.e. for $d_p/D_b < 0.1$, $\Phi_{pc}$ does not depend on particle size and is equal to $\Phi_{pc} \approx 0.03$, which corresponds to an increase by 40% of the particle strengthening with respect to previous work [26]. Note however that the 30 μm particles induce a slightly higher $\Phi_{pc}$ value, which means that the optimal strengthening is in fact limited to a size ratio $d_p/D_b$ smaller than 0.06. This suggests that previous work did not provide such an optimal strengthening because the investigated particle-to-bubble size ratio was too large, i.e. $d_p/D_b > 0.3$. Figure 6 also shows that the rigidity percolation model fails to describe our data in the range $d_p/D_b < 0.1$, which means that an alternative modeling should be proposed.

For such small particle-to-bubble size ratios, images of the material microstructure reveal a specific organization of the particles within the foam pore space. The strong increase measured for the elastic modulus is clearly associated to the formation of a thick skeleton made of packed particles. Note that for low particle contents (few percents), the contribution of the particle skeleton exceeds by far the capillary effects due to bubbles and it is therefore necessary to describe explicitly the elasticity of the skeleton. The simplest modeling for this contribution is based on the elastic behavior of solid foams, as proposed by Gibson & Ashby [30]. The shear modulus due to a skeleton of solid and continuous matrix can be expressed as $G_s \approx aG_0(1-\phi)^2$, where $G_0$ is the bulk shear modulus of the matrix forming the skeleton and $a \approx 1$ is a numerical coefficient. Note that for the system investigated in the present study, the elasticity of the bubbles also has to be accounted for, i.e. $G_b \equiv G'(\phi, 0) \approx b(\gamma/D_b)\phi(\phi - \phi_c)$. Focusing on situations corresponding to high enough $\phi_p$ values for the skeleton to be formed, we expect that $G_s \gg G_b$ and coupling effects can be therefore ignored. For such situations, the global elastic behavior can be obtained by summing those two contributions:

$$G' \approx G_0(1-\phi)^2 + b(\gamma/D_b)\phi(\phi - \phi_c) \quad \text{(eq. 1)}$$

As the skeleton is made of packed particles, its volume fraction is given by $1 - \phi \approx \phi_p/\phi_c$. Therefore, the reduced elastic modulus is:

$$\frac{G'(\phi, \phi_p)}{G'(\phi, 0)} = 1 + \frac{G_0}{b(\gamma/D_b)} f(\phi_p) \quad \text{(eq. 2)}$$

where $f(\phi_p) = [(\phi_c/\phi_p - 1)(\phi_c/\phi_p - 1 - \phi_c^2/\phi_p)]^{-1}$. $G_0$ is the only fitting parameter and it represents the shear elastic modulus of the bulk granular packing forming the skeleton. Using $G_0$ = 150 kPa in eq. 2, good agreement is obtained with our data over the full range of $\phi_p$ values (see figure 3). Note however that eq. 2 is expected to apply as the granular skeleton is formed, i.e. $\phi_p > 0.03$. $G_0$ value is surprisingly high with regard to both involved bubble capillary pressure, $4\gamma/D_b \approx 300$ Pa, and hydrostatic pressure, $\rho g \bar{H} \sim 400$ Pa. In the following we discuss about the origin of apparent rigidity of the studied foams.

The key understanding element is presented figure 5b: even a small confinement pressure induces high values for the elastic modulus of granular matter [31,32]. Therefore, it is highly conceivable that in those foamy systems such a small confinement pressure is provided by bubbles interface pushing on packed particles. Indeed, because the liquid is free to drain within the packed particles, this confinement pressure $P_C$ is set by hydrostatic pressure and it is therefore equal to $P_c \approx \rho g \overline{H} \sim 400$ Pa. In assuming such a confinement pressure, the ratio $G_0/P_C \approx 400$, which falls slightly below values measured for the bulk granular material at $\varepsilon \approx 10^{-3}$ (see figure 5b). Note that for high strain amplitudes the behavior of the particle-free foam is recovered, which indicates that bubble confinement effect is lost due to the yielding of the bubble assembly. In order to go deeper in the comparison with the bulk granular material, one has to account for localization effects. For a macroscopic deformation $\varepsilon$ applied on the foam, elastic energy $G'\varepsilon^2$ can also be written in terms of the local deformation $\varepsilon_\ell$ that characterizes the granular skeleton: $G_0 \varepsilon_\ell^2 \, \phi_p/\phi_c$. Using the Gibson and Ashby relation for $G'$, i.e. $G' \simeq G_0 (\phi_p/\phi_c)^2$, the local deformation is related to the macroscopic deformation by $\varepsilon_\ell \approx \varepsilon (\phi_p/\phi_c)^{1/2}$, i.e. $\varepsilon_\ell \approx \varepsilon/4$ for the particle volume fractions corresponding to the granular skeleton. Although rather weak, localization effect is taken into account and curves corresponding to the granular material are plotted in figures 2a (inset) and 2b. This clearly shows that as the particle volume fraction increases, the critical deformation of particle-loaded foams decreases and gets closer to the critical deformation of granular matter (figure 2b). The granular solid-like behavior also manifests in the global evolution of the curve $G'(\varepsilon)$ : the two-step yielding of particle-loaded foams can be analyzed as the yielding of the granular material occurring at small strains (see inset in figure 2a) and the further yielding of the bubble assembly occurring at larger strain. Note also that the high values measured for the loss modulus at small strain amplitudes also account for the yielding of the granular skeleton.

## 5. Conclusion

We have studied the rheological behavior of particle-loaded liquid foams characterized by particle-to-bubble size ratios smaller than 0.1. With respect to previous work, our results highlight a critical size ratio below which a new elastic regime, characterized by unequaled modulus values as well as independence of size ratio, has been identified. This behavior, which has been clearly correlated to the formation of a thick granular skeleton of tightly packed particles, is not correctly described by the so-called rigidity percolation model [26]. We have shown that this increased elasticity originates from the strong synergistic effects that can be brought about in those particle-loaded foams: the particles are small enough for forming an interstitial granular skeleton and the bubbles interface imposes a small but crucial confinement pressure on the packed particles, therefore providing significant mechanical strength to the granular skeleton. In spite of the small volume fraction occupied by the granular skeleton, its effect on the mechanical behaviour is however significant. We have shown that a simple model accounting for both the intrinsic elastic modulus of the bulk granular material and the volume fraction of the skeleton reproduces well the effect of particle loading.

The investigated system is characterized by a rather small confinement pressure exerted by the foam bubbles and one can expect that a stronger effect could be obtained by increasing this parameter. Figure 5 suggests that an increase by a factor 10 could be obtained for the elastic modulus of the bulk granular skeleton, i.e. $G_0 \approx 2$ MPa. In practice, such an increase could be performed by reducing the amount of liquid available in the foam. Such conditions could be explored in the future.

The approach we propose is expected to be valid as long as particles are large enough to be excluded from the foam films and to form the skeleton morphology described above. Note however that incorporating particles into the foam films would imply particles size within the colloidal scale, i.e. a few hundred nanometers. Moreover, colloidal interactions would modify the mechanical behavior of particle packings, providing intrinsic elastic properties, i.e. without need of bubble confinement pressure.

**Acknowledgements**

We thank D. Hautemayou and C. Mézière for technical support. We gratefully acknowledge financial support from Agence Nationale de la Recherche (Grant no. ANR-13-RMNP-0003-01) and French Space Agency (convention CNES/70980). We thank X. Chateau and G. Ovarlez for fruitful discussions.

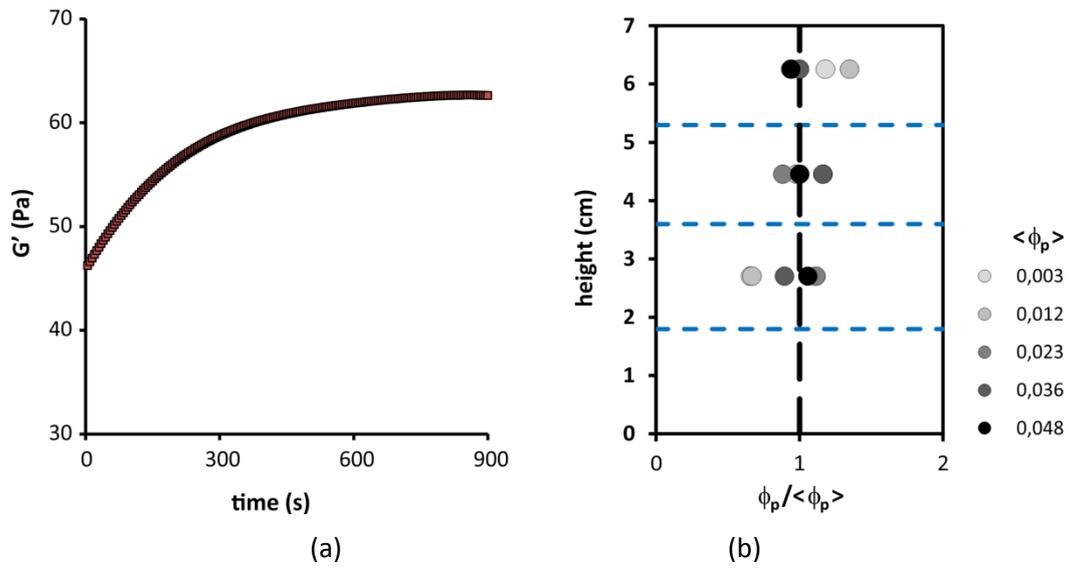

Fig.1. (a) Typical time evolution of the shear elastic modulus ($\varepsilon = 10^{-3}$; 1Hz) for a particle-loaded foam, straight after the filling of the cell. This evolution is due to drainage process and the value of modulus reported in the following is the value measured at long times. (b) Example vertical profile for the particle (20 µm) volume fraction $\phi_p$ measured in drained particle-loaded foams, for several average values $\langle \phi_p \rangle$.

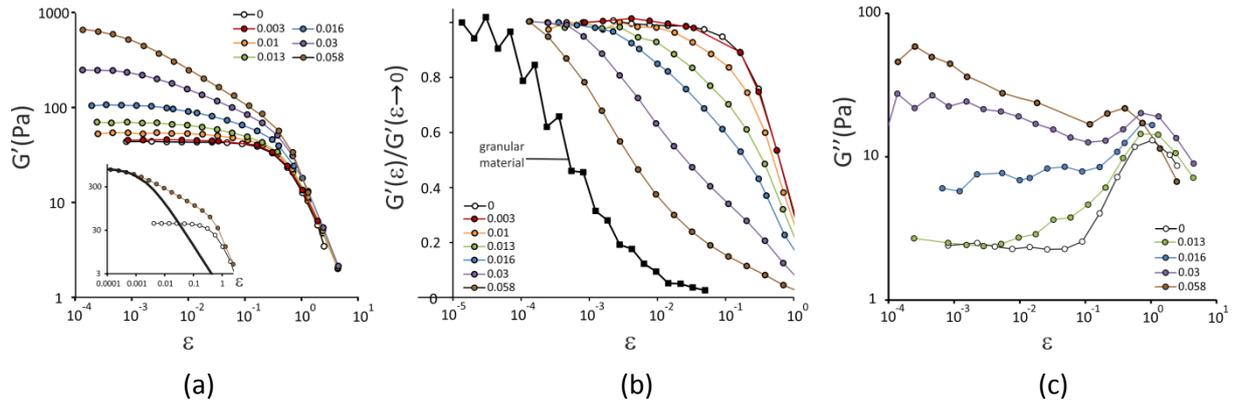

Fig. 2. Oscillatory rheology (1 Hz) of liquid foams laden with 20μm particles for several particle volume fractions. a) Shear elastic modulus $G'(\varepsilon)$ as a function of strain amplitude ; inset: illustration of the two-step yielding of particle-loaded foams. The continuous line represents the behavior of the bulk granular skeleton, estimated by the function $G_0(1-\phi)^2 \, G'(\varepsilon)/G'(\varepsilon \approx 0)$, where the curve $G'(\varepsilon)/G'(\varepsilon \approx 0)$ for the bulk granular material is presented in figure 2b (black squares), and $G_0(1-\phi)^2 \approx 800$ Pa represents the contribution of the granular skeleton at small strain amplitudes. b) Reduced elastic modulus $G'(\varepsilon)/G'(\varepsilon \to 0)$ as a function of strain amplitude – the black squares correspond to data obtained for the bulk granular material and presented in figure 5a. Note that localization effects have been accounted for (see the main text for details). c) Shear viscous modulus $G''(\varepsilon)$ as a function of strain amplitude.

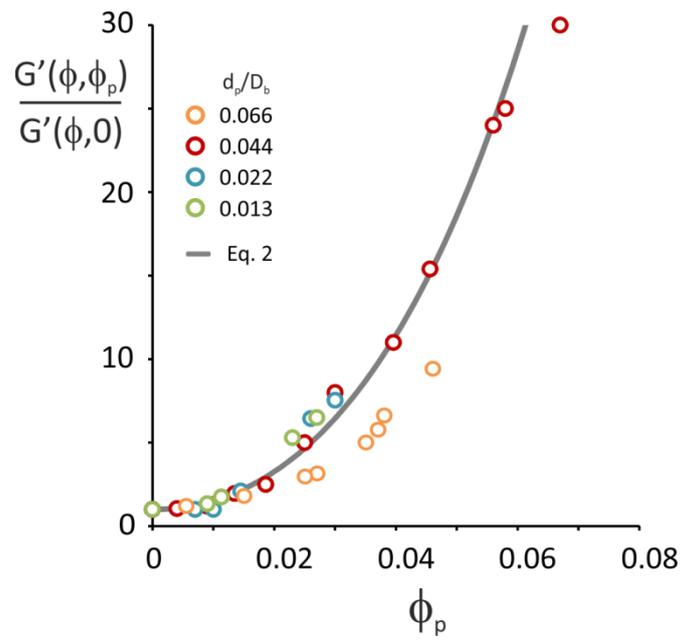

Fig. 3: Reduced elastic modulus $G'(\phi, \phi_p)/G'(\phi, 0)$ as a function of the particle volume fraction for several particle-to-bubble size ratios. The grey line corresponds to the modeling presented in eq. 2.

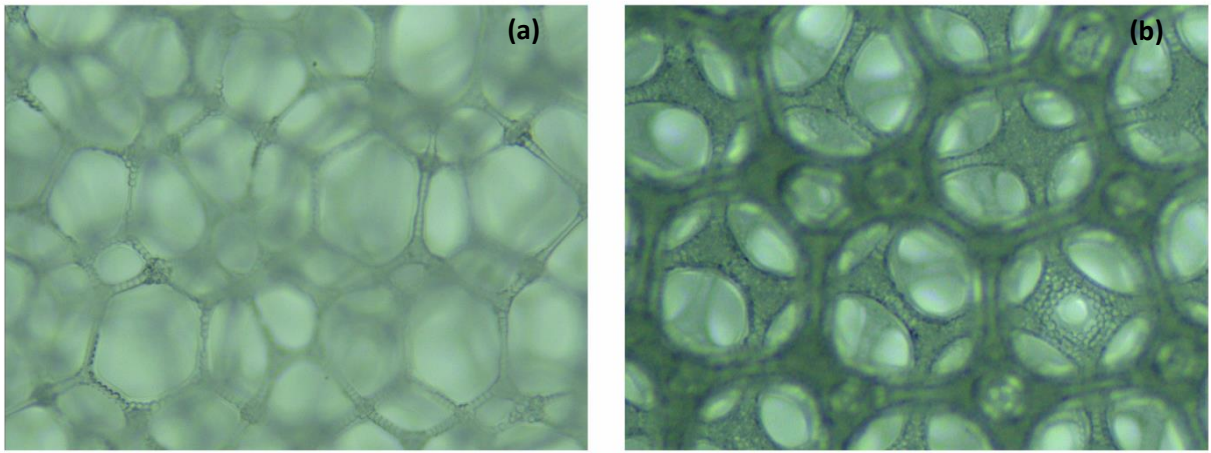

Fig. 4: 20µm particle loaded foam. The filling of the foam network by the polystyrene beads is observed for two $\phi_p$ values: (a) 0.01 and (b) 0.04. Note the enlargement of the particle skeleton as $\phi_p$ increases.

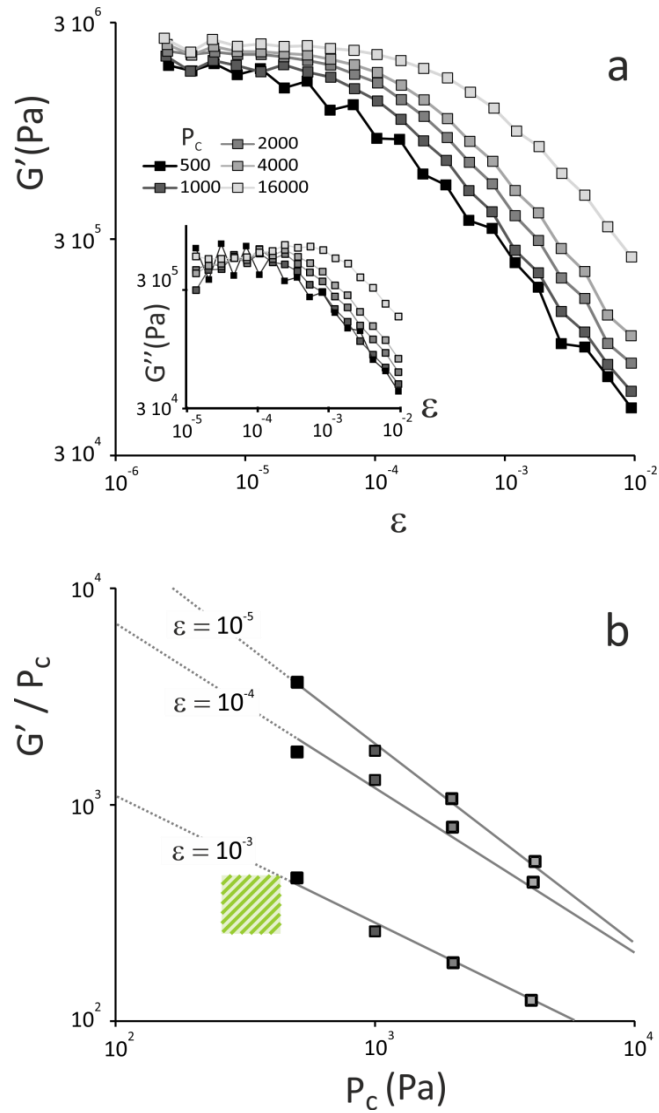

Fig. 5: (a) Shear elastic modulus of the bulk granular material as a function of strain amplitude for several confinement pressures $P_C$. Inset: Shear loss modulus of the bulk granular material as a function of strain amplitude for several confinement pressures $P_C$. (b) Ratio of the measured elastic modulus to the confinement pressure applied to the granular material as a function of the confinement pressure for several strain values. The shaded area represents the behavior of particle-loaded foams.

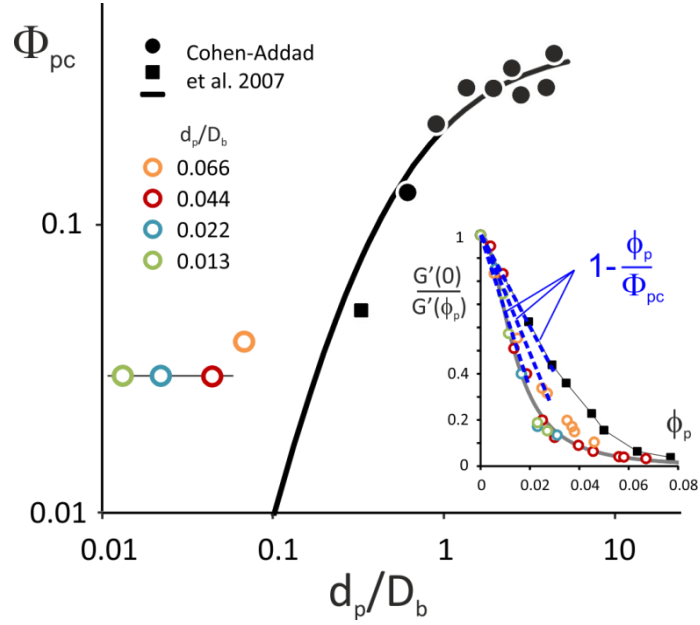

Fig. 6 : Rigidity percolation threshold $\Phi_{pc}$ as introduced by [26]. Black symbols represent the data from [26] and the empty symbols represent our results. Inset: The relation $G'(\phi_p)/G'(0) \approx 1/(1 - \phi_p/\Phi_{pc})$ is used to determine $\Phi_{pc}$, as shown by the three dashed lines. The grey line represents eq. 2. Main figure: $\Phi_{pc}$ as a function of the particle-to-bubble size ratio. The thick black solid line represents the theoretical percolation threshold [26]: $\Phi_{pc} = \Phi_{pc}^{eff}(1 + 2h/d_p)^{-3}$, with $\Phi_{pc}^{eff} = 0.42$ and $h = 3.8$ μm.